# Increasing Average Fidelity by Using Non-Maximally Entangled Resource in Teleportation of Superposed Coherent States

Hari Prakash[1] and Manoj K Mishra[2]

[1]Department of physics, University of Allahabad, Allahabad, INDIA

[2]Department of physics, University of Allahabad, Allahabad, INDIA

[1]prakash_hari123@rediffmail.com, [2]manoj.qit@gmail.com

**Abstract:** We consider the problem of teleporting superposed coherent state using non-maximally entangled coherent state as quantum channel and study the effect of entanglement over quality of teleportation. We show that if maximally entangled coherent state (MECS) used by Van Enk and Hirota [Phys. Rev. A 64, 022313 (2001)] and by H Prakash et al. [Phys. Rev. A 75, 044305 (2007)] is replaced by a particular non-maximally entangled coherent state (NMECS), average fidelity of quantum teleportation increases appreciably at small coherent amplitudes. Since it is very challenging to produce superposed coherent states of large coherent amplitudes, the particular NMECS appear to be a good quantum channel for teleportation at small coherent amplitudes for practical implementation.

First idea of quantum teleportation is due to Bennett et al [1] which involves complete transfer of an unknown quantum state of a system to another system across space using quantum entanglement also referred as Einstein-Podolsky-Rosen correlations [2]. Experimental teleportation of polarized single photon state has been realized using standard bi-photonic entangled states (The Bell state) [3-5] as quantum channel. The experiments using standard bi-photonic entangled states has been successful in proving the principle of quantum teleportation but are commercially inapplicable due to low efficiency in production and detection of single photons and decoherence due to photon absorption.

In recent past another form of entangled state-- an entangled coherent state (ECS) [6] has been attracted much attention. Gerry [7] proposed nonlinear Mach-Zehnder interferometer as a device to transform a pair of coherent states into an ECS. Howell and Yeazell [8] proposed generation of ECS via two non-demodulation measurements. Recently Liao and Kuang [9] given a scheme to generate ECS of two micro-cavity fields coupling to a SQUID- based cooper pair box charge qubit. Some authors [10-13] studied entanglement properties of ECS and also devised methods for preparation of a large class of ECS which are maximally entangled. Hirota and Sasaki [13], has been shown that ECS are stronger against decoherence due to photon absorption than standard bi-photonic entangled states. Large numbers of schemes has been proposed for generating superposed coherent states (SCS) [14-21] which in turn attracted researchers to use them as qubit in different quantum information processing tasks. One advantage of using SCS as qubit is that, this reduces the necessity of practical "exact" single photon sources which are yet unavailable. Van Enk and Hirota [2] have shown how to teleport a SCS using ECS with success probability equal to ½ and Wang [23] presented a very similar scheme for teleporting bipartite ECS with success probability equal to ½. Further H. Prakash et al [24] by amending the photon counting scheme (The Bell measurement scheme) reported almost perfect teleportation with success probability almost equal to unity. Liao and Kuang [25], Phien and Nguyen [26] and the authors [27] studied teleportation of four-component bipartite ECS. All of the above schemes used maximally ECS as quantum channel. One important motivation behind the work on teleportation is the possibility of implementing quantum computation using SCS and ECS [28, 29]. But this does not imply that all above given schemes for teleporting quantum information encoded in SCS-qubit using ECS are experimentally realizable with currently available technology. Although SCS and ECS are stronger against decoherence due to photon absorption [13] than polarized single photon state and standard bi-photonic entangled states respectively, but are still sensitive to decoherence.

In present paper we will consider problem of teleporting SCS via an arbitrary bipartite ECS of the form,

$$|E\rangle_{1,2} = N(\cos\tfrac{\theta}{2}|\alpha,\alpha\rangle + \sin\tfrac{\theta}{2}e^{i\varphi}|-\alpha,-\alpha\rangle), \quad \{\theta \in [0,\pi], \varphi \in [0,2\pi)\} \tag{1}$$

where,

$$N = (1 + x^4 \sin\theta \cos\varphi)^{-1/2} \tag{2}$$

and $x \equiv \exp(-|\alpha|^2)$. We first briefly review entanglement properties of above given ECS. Coherent states $|\pm\alpha\rangle$, can be written as linear superposition of state with even numbers of photons $|+\rangle$ and state with odd numbers of photons $|-\rangle$ given by,

$$|\pm\alpha\rangle = \tfrac{1}{\sqrt{2}}[(1+x^2)^{1/2}|+\rangle \pm (1-x^2)^{1/2}|-\rangle], \tag{3}$$

where,

$$|\pm\rangle = [2(1\pm x^2)]^{-1/2}(|\alpha\rangle \pm |-\alpha\rangle). \tag{4}$$

Using (3) for $|\pm\alpha\rangle$, ECS (1) becomes,

$$|E\rangle_{1,2} = N[\tfrac{1}{2}\{C_+(1+x^2)|+,+\rangle + C_-(1-x^4)^{-1/2}(|+,-\rangle + |-,+\rangle) + C_+(1-x^2)|-,-\rangle\}] \tag{5}$$

where,

$$C_\pm = \cos\tfrac{\theta}{2} \pm \sin\tfrac{\theta}{2}e^{i\varphi}. \tag{6}$$

Concurrence of above ECS using relation, $C = |\langle \psi | \tilde{\psi} \rangle|$ with $|\tilde{\psi}\rangle = \sigma_y |\psi^*\rangle$ given by Wooters [30] is,

$$C = 2(|C_+|^2 - |C_-|^2) = (1 - x^4)\sin\theta(1 + x^4 \sin\theta\cos\varphi)^{-1}. \tag{7}$$

It is clear that for $\theta = \pi/2$ and $\varphi = \pi$, C=1 i.e., $|E\rangle_{1,2}$ reduces to a maximally entangled coherent state (MECS) that was used in Ref. [22, 23, 24]. For $\theta = \pi/2$ and $\varphi = 0, 2\pi$, $|E\rangle_{1,2}$ is non-maximally entangled coherent state (NMECS) with $C = (1 - x^4)(1 + x^4)^{-1}$ which becomes almost equal to unity for large coherent amplitude. In other words, it remains non-maximally entangled only for low coherent amplitudes. Thus, now we will consider the problem of teleporting SCS of the form,

$$|I\rangle_0 = \varepsilon_+ |\alpha\rangle + \varepsilon_- |-\alpha\rangle = A_+ |+\rangle + A_- |-\rangle, \tag{8}$$

with normalization conditions, $[|\varepsilon_+|^2 + |\varepsilon_-|^2 + 2x^2 \text{Re}(\varepsilon_+^* \varepsilon_-)] = 1$ and $|A_+|^2 + |A_-|^2 = 1$, with the interrelationship between coefficients $A_\pm$ and $\varepsilon_\pm$ given by,

$$A_\pm = (\varepsilon_+ \pm \varepsilon_-)[\tfrac{1}{2}(1 \pm x^2)]^{1/2}, \varepsilon_\pm = \tfrac{1}{\sqrt{2}}[A_+(1+x^2)^{-1/2} \pm A_-(1-x^2)^{-1/2}], \tag{9}$$

using ECS (1) and then investigate the effect of entanglement over minimum average fidelity (defined later in this paper) by varying the entanglement parameters $\theta, \varphi$.

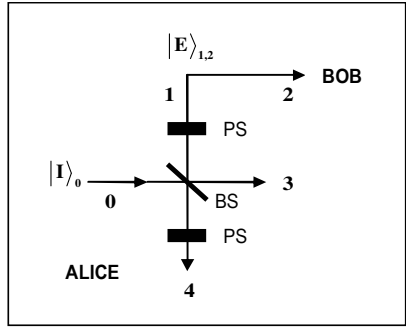

Fig. 1. Scheme for teleportation. PS represents – $\pi/2$ phase shifters that convert state $|\alpha\rangle$ to $|-i\alpha\rangle$. BS represents symmetric beam splitters. Bold numbers denote modes.

The initial state of the system is $|\psi\rangle_{0,1,2} = |I\rangle_0 |E\rangle_{1,2}$. Our teleportation scheme is similar to that used by Van Enk and Hirota [22] and H. Prakash et al [24] and is shown in Fig. 1. Modes 0 and 1 are with Alice (sender) and mode 2 is with Bob (receiver). After passing the quantum modes 0, 1, and 2 through the scheme shown in Fig. 1, the final output state is found to be,

$$|\psi\rangle_{3,4,2} = N[\varepsilon_+ \{\cos\tfrac{\theta}{2} |\sqrt{2}\alpha, 0, \alpha\rangle + \sin\tfrac{\theta}{2} e^{i\varphi} |0, \sqrt{2}\alpha, -\alpha\rangle\}$$
$$+ \varepsilon_- \{\cos\tfrac{\theta}{2} |0, -\sqrt{2}\alpha, \alpha\rangle + \sin\tfrac{\theta}{2} e^{i\varphi} |-\sqrt{2}\alpha, 0, -\alpha\rangle\}]_{3,4,2} \tag{10}$$

Alice performs Photon counting in modes 3 and 4, and as is clear from (10) one of the counts is always zero. We follow Prakash et al [24] and expand coherent states $|\pm\sqrt{2}\alpha\rangle$ which are with Alice into no-photon state (the vacuum state $|0\rangle$), states with nonzero even numbers of photons ($|\text{NZE}, \sqrt{2}\alpha\rangle$) and states with odd numbers of photons ($|\text{ODD}, \sqrt{2}\alpha\rangle$) defined by these authors and write [24],

$$|\pm\sqrt{2}\alpha\rangle = x|0\rangle + \tfrac{1}{\sqrt{2}}(1-x^2)|\text{NZE},\sqrt{2}\alpha\rangle \pm [\tfrac{1}{2}(1-x^4)]^{1/2}|\text{ODD},\sqrt{2}\alpha\rangle, \qquad (11)$$

But we expand states $|\pm\alpha\rangle$ with Bob using (3). The final output state is then found in the form,

$$\begin{aligned}
|\psi\rangle_{3,4,2} = \tfrac{N}{\sqrt{2}}\big[&|0,0\rangle_{3,4}\, x(\varepsilon_+ + \varepsilon_-)\{p^{-1}C_+|+\rangle + q^{-1}C_-|-\rangle\}_2 \\
&+ \tfrac{1}{2}\{q^{-2}|\text{NZE},0\rangle_{3,4}\{p^{-1}(C_+A_+p + C_-A_-q)|+\rangle + q^{-1}(C_-A_+p + C_+A_-q)|-\rangle\}_2 \\
&+ q^{-2}|0,\text{NZE}\rangle_{3,4}\{p^{-1}(C_+A_+p - C_-A_-q)|+\rangle + q^{-1}(C_-A_+p - C_+A_-\ q)|-\rangle\}_2 \\
&+ (pq)^{-1}|\text{ODD},0\rangle_{3,4}\{p^{-1}(C_-A_+p + C_+A_-q)|+\rangle + q^{-1}(C_+A_+p + C_-A_-q)|-\rangle\}_2 \\
&+ (pq)^{-1}|0,\text{ODD}\rangle_{3,4}\{p^{-1}(C_-A_+p - C_+A_-q)|+\rangle + q^{-1}(C_+A_+p - C_-A_-q)|-\rangle\}_2 \big]
\end{aligned} \qquad (12)$$

where, $p \equiv (1+x^2)^{-1/2}$ and $q \equiv (1-x^2)^{-1/2}$.

From (12), we see that there are five possible photon counting results, (0, 0), (0, NZE), (NZE, 0), (ODD, 0) and (0, ODD). Alice conveys these photon counting results to Bob via a two bit classical channel, on the basis of which Bob performs required unitary operation in mode 2 to get a faithful replica of the information state. Since state with Bob depends on arbitrary entanglement parameters ($\theta, \varphi$), we will use two different strategies to recover replica of information state to as large fidelity as possible:

*Strategy 1*: For $|C_+| \geqslant |C_-|$ (i.e., for $\cos\varphi \geqslant 0$), Bob performs following unitary operations:

$$U_{0,0} = U_{\text{NZE},0} = I,\ U_{0,\text{NZE}} = |+\rangle\langle+|-|-\rangle\langle-|,\ U_{\text{ODD},0} = |+\rangle\langle-|+|-\rangle\langle+|,\ U_{0,\text{ODD}} = |+\rangle\langle-|-|-\rangle\langle+|. \quad (13)$$

Teleported states with Bob after above given unitary operations are,

$$\begin{aligned}
|T_{0,0}\rangle &= \{C_+p^{-1}|+\rangle + C_-q^{-1}|-\rangle\}_2 \\
|T_{\text{NZE},0}\rangle &= \{(C_+A_+p + C_-A_-q)p^{-1}|+\rangle + (C_+A_-q + C_-A_+p)q^{-1}|-\rangle\}_2 \\
|T_{0,\text{NZE}}\rangle &= \{(C_+A_+p - C_-A_-q)p^{-1}|+\rangle + (C_+A_-q - C_-A_+p)q^{-1}|-\rangle\}_2 \\
|T_{\text{ODD},0}\rangle &= \{(C_+A_+p + C_-A_-q)q^{-1}|+\rangle + (C_+A_-q + C_-A_+p)p^{-1}|-\rangle\}_2 \\
|T_{0,\text{ODD}}\rangle &= \{(C_+A_+p - C_-A_-q)q^{-1}|+\rangle + (C_+A_-q - C_-A_+p)p^{-1}|-\rangle\}_2
\end{aligned} \qquad (14)$$

Fidelity is defined as the overlap of the teleported state ($|T\rangle$) over original information state ($|I\rangle$) given by, $F = |\langle I|T\rangle|^2$. Average fidelity is defined as the summation of products of probability of occurrence and fidelity for all possible photon counting results given by $F_{av.} = \sum F_i P_i$, where $P_i$ and $F_i$ stands for probability of occurrence and fidelity of the teleported state for $i^{th}$ photon counting result respectively. Putting $A_+ = \cos\frac{\omega}{2}$ and $A_- = \sin\frac{\omega}{2} e^{i\xi}$, average fidelity is,

$$\begin{aligned}
F_{av.} = \frac{1}{4} N^2 [ & 2x^2(1+\cos\omega)(1+x^2)^{-1}\{1 + x^2 \sin\theta\cos\varphi + \cos\omega(\sin\theta\cos\varphi + x^2) \\
& + (1-x^4)^{1/2} \sin\omega(\cos\theta\cos\xi - \sin\theta\sin\varphi\sin\xi)\} + (1+\sin\theta\cos\varphi)(1 - x^2\cos\omega)^2 \\
& + (1-x^2)^2 \{1 + \sin\theta\cos\varphi + (1-x^4)^{-1} \sin^2\omega(1-\sin\theta\cos\varphi)(\cos^2\xi + x^4 \sin^2\xi)\} \\
& + (1-x^4)(1-\sin\theta\cos\varphi)\sin^2\omega\cos^2\xi]
\end{aligned} \quad (15)$$

*Strategy 2:* For $|C_+| < |C_-|$ (i.e., for $\cos\varphi < 0$), Bob performs following unitary operations:

$$U_{0,0} = U_{ODD,0} = I, U_{0,ODD} = |+\rangle\langle+| - |-\rangle\langle-|, U_{NZE,0} = |+\rangle\langle-| + |-\rangle\langle+|, U_{0,NZE} = |+\rangle\langle-| - |-\rangle\langle+|. \quad (16)$$

Teleported states with Bob after above given unitary transformations are,

$$\begin{aligned}
|T_{0,0}\rangle &= \{C_+ p^{-1}|+\rangle + C_- q^{-1}|-\rangle\}_2 \\
|T_{NZE,0}\rangle &= \{(C_- A_+ p + C_+ A_- q)q^{-1}|+\rangle + (C_- A_- q + C_+ A_+ p)p^{-1}|-\rangle\}_2 \\
|T_{0,NZE}\rangle &= \{(C_- A_+ p - C_+ A_- q)q^{-1}|+\rangle + (C_- A_- q - C_+ A_+ p)p^{-1}|-\rangle\}_2 \\
|T_{ODD,0}\rangle &= \{(C_- A_+ p + C_+ A_- q)p^{-1}|+\rangle + (C_- A_- q + C_+ A_+ p)q^{-1}|-\rangle\}_2 \\
|T_{0,ODD}\rangle &= \{(C_- A_+ p - C_+ A_- q)p^{-1}|+\rangle + (C_- A_- q - C_+ A_+ p)q^{-1}|-\rangle\}_2
\end{aligned} \quad (17)$$

Following previous strategy, average fidelity for this is,

$$\begin{aligned}
F_{av.} = \frac{1}{4} N^2 [ & 2x^2(1+\cos\omega)(1+x^2)^{-1}\{1 + x^2 \sin\theta\cos\varphi + \cos\omega(\sin\theta\cos\varphi + x^2) \\
& + (1-x^4)^{1/2} \sin\omega(\cos\theta\cos\xi - \sin\theta\sin\varphi\sin\xi)\} + (1-x^2)^2 \{(1-x^4)^{-1}(1-\sin\theta\cos\varphi) \\
& \times (1-x^2\cos\omega)^2 + (1-\sin\theta\cos\varphi)\sin^2\omega\cos^2\xi\} + (1-x^4)(1-\sin\theta\cos\varphi) \\
& + \sin^2\omega(1+\sin\theta\cos\varphi)(\cos^2\xi + x^4 \sin^2\xi)\}]
\end{aligned} \quad (18)$$

To investigate the effect of entanglement over quality of teleportation, we will plot minimum average fidelity, $F_{min., av.}$ (defined as minimum of average fidelity ($F_{av.}$ in Eq. 17, 18) over all possible information states i.e., over angles $\omega, \xi$) with respect to entanglement parameters $\theta, \varphi$. Fig. 2 shows this for different values of mean photon number ($|\alpha|^2$). From Fig.2 it is clear that for any given ECS, $F_{min., av.}$ always increases with increase in $|\alpha|^2$. Interesting result is that at low $|\alpha|^2$, the values of $F_{min., av.}$ for the two maxima at $\theta = \pi/2, \varphi = 0 \& 2\pi$ (which corresponds to NMECS) is higher than that at $\theta = \pi/2, \varphi = \pi$ (corresponds to MECS used in Ref. [22-24]). At higher values of $|\alpha|^2$, for both cases $F_{min., av.}$ becomes comparable and almost equal to unity which is expected since NMECS in context becomes almost MECS at higher values of $|\alpha|^2$ as mentioned earlier (Eq.7). Substituting $\theta = \pi/2, \varphi = 0 \,\&\, 2\pi$ (ECS of Eq.(1) is a special NMECS) in (15), expression for average fidelity is becomes $F_{av.}^{(1)} = 1 - [x^2(1+x^2)\sin^2\omega][2(1+x^4)]^{-1}$. The minimum value of $F_{av.}$ occurs at $\omega = \pi/2$ i.e., when $|A_+| = |A_-| = 1/\sqrt{2}$ and is given by,

$$F_{min.,av}^{(1)} = 1 - [x^2(1+x^2)][2(1+x^4)]^{-1}. \tag{19}$$

Substituting $\theta = \pi/2, \varphi = \pi$ (ECS of Eq. (1) is then MECS and the same as used in references [22-24]) in (18), expression for average fidelity is $F_{av.}^{(2)} = 1 - 2x^2\cos^2\frac{\omega}{2}(\cos^2\frac{\omega}{2} + x^2\sin^2\frac{\omega}{2})(1+x^2)^{-2}$. The minimum value of $F_{av.}$ occurs at $\omega = 0$ i.e., when $|A_+| = 1, |A_-| = 0$ and is given by

$$F_{min.,av}^{(2)} = 1 - 2x^2(1+x^2)^{-2}. \tag{20}$$

Difference between the two minimum average fidelities is given as,

$$D = F_{av.min.}^{(1)} - F_{av.min.}^{(2)} = [x^2(3+x^4)(1-x^2)][2(1+x^4)(1+x^2)^2]^{-1} \tag{21}$$

Variation of difference D with |α|² is shown in Fig. 2(d). The maximum difference is ≈ 0.17 at |α|²≈0.6.

Variation of $F_{min.,av.}$ with $\theta$ and $\varphi$ is shown in Fig. 2(a)-(c) for $|\alpha|^2$=0.5, 1.0 and 1.5. it is seen clearly that $F_{min.,av.}$ is peaked at $(\theta = \pi/2)$ and $(\varphi = 0, \pi \,\&\, 2\pi)$. Peak at $\varphi = 0$ (same as $\varphi = 2\pi$) refers to non-maximally ECS $\sim (|\alpha,\alpha\rangle + |-\alpha,-\alpha\rangle)_{a,b}$ while the peak at $\varphi = \pi/2$ refers to the maximally ECS used in references [22-24]. It is clear from Fig. 2(a)-(c) that the NMECS gives a high peak than the MECS. The maxima for $F_{min.,av.}$, obtained for $\theta = \pi/2, \varphi = 0, \pi \,\&\, 2\pi$, gives ECS of the form, $N(|\alpha,\alpha\rangle \pm |-\alpha,-\alpha\rangle)_{a,b}$ where N= $[2(1\pm x^2)]^{-1/2}$. These can be deterministically generated by illuminating a 50:50 beam splitter with SCS of the form, $N(|\beta\rangle \pm |-\beta\rangle)$ where $\beta = \sqrt{2}\alpha$ with mean photon number |β|². It is known that SCS can be generated from a coherent state by nonlinear interaction in a Kerr media [14-15], but presently available Kerr nonlinearity is too small [7, 15]. For this reason schemes using very small Kerr nonlinearity [16] and without Kerr nonlinearity [17-19] have been proposed. Alexi [20] and Nielsen [21] have experimentally demonstrated the generation of small SCS with |β|² ≈1 by subtracting one photon from a squeezed vacuum. These kittens can be used to generate ECS with $|\alpha|^2$≈0.5. For small coherent amplitude, such as those with $|\alpha|^2$≈0.5, it is clear that use of NMECS with the strategies given in the present paper will give appreciably higher average fidelity than the earlier schemes [22-24].

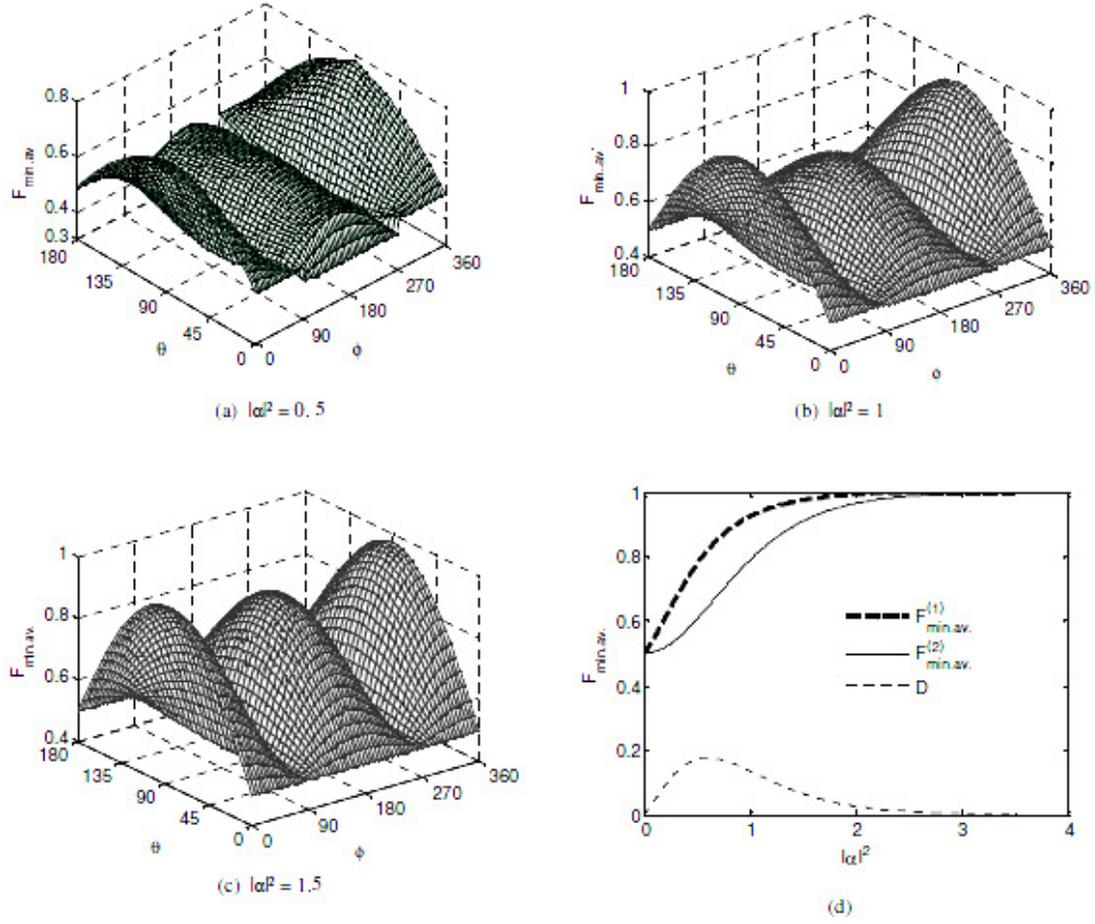

Fig. 2. Plots (a), (b) and (c) shows variation of $F_{min.av}$ with respect to entanglement parameters $\theta$ and $\Phi$, for different values of mean photon number $|\alpha|^2$. Plot (d) shows variation of $F^{(1)}_{min.av.}$ for non-maximally ECS, $F^{(2)}_{min.av.}$ for maximally ECS and difference D with respect to mean photon number $|\alpha|^2$.

Acknowledgement:

We are grateful to Prof. N. Chandra and Prof. R. Prakash for their interest and stimulating discussions. Discussions with Ajay Kumar Yadav, Ajay Kumar Maurya and Vikram Verma are gratefully acknowledged. One of the authors (MKM) acknowledges the UGC for financial support under UGC-SRF fellowship scheme.